\documentclass[preprint,aps,draftams,math,amssymb,superscriptaddress,unsortedaddress]{revtex4}
\usepackage{graphicx}
\usepackage{dcolumn}
\usepackage{bm}
\usepackage[usenames,dvips]{color}
\usepackage{gensymb}
\usepackage{amsmath}
\begin{document}

\title{Effect of magnetic disorder and strong electron correlations on the thermodynamics of CrN}

\author{B. Alling}
\email{bjoal@ifm.liu.se}
\affiliation{Department of Physics, Chemistry and Biology (IFM), \\
Link\"oping University, SE-581 83 Link\"oping, Sweden.}
\author{T. Marten}
\affiliation{Department of Physics, Chemistry and Biology (IFM), \\
Link\"oping University, SE-581 83 Link\"oping, Sweden.}
\author{I.~A. Abrikosov}
\affiliation{Department of Physics, Chemistry and Biology (IFM), \\
Link\"oping University, SE-581 83 Link\"oping, Sweden.}
\date{\today}

\begin{abstract}
We use first-principles calculations to study the effect of magnetic disorder and electron correlations on the structural and thermodynamic properties of CrN. We illustrate the usability of a special quasirandom structure supercell treatment of the magnetic disorder by comparing with coherent potential approximation calculations and with a complementary magnetic sampling method. The need of a treatment of electron correlations effects beyond the local density approximation is proven by a comparison of LDA+U calculations of structural and electronic properties with experimental results. When magnetic disorder and strong electron correlations are taken into account simultaneously, pressure and temperature induced structural and magnetic transitions in CrN can be understood. 

\end{abstract}

\maketitle

\section{Introduction}
Transition metal nitrides have attracted much interest due to their excellent performance in a long range of industrial application such as hard protective coatings on cutting tools, diffusion barriers, and wear resistant electrical contacts. CrN is not as hard as for instance TiN~\cite{Holleck1986}, but it is superior to TiN in solving large concentrations of AlN in the rock salt phase giving rise to the Cr$_{1-x}$Al$_{x}$N solid solutions highly valued in hard coatings applications~\cite{Knotek1990,Wuhrer2004,Reiter2005}. Furthermore, CrN on its own can be found in coating application for metal forming and plastic moulding purposes~\cite{Vetter1995, Persson2001}. From a fundamental physics point of view the study of CrN has provided new insights but also raised questions about magneto-driven structural transitions. It is known that a magnetic order-disorder transition at temperatures around 280 K is associated with an orthorhombic to cubic structural transition~\cite{Corliss1960}, recently observed to be reversible under small pressures~\cite{Rivadulla2009}. On the other hand in epitaxially stabilized cubic thin films no sign of magnetic ordering has been seen~\cite{Gall2002, Sanjines2002}. On the theoretical side the concept of magnetic stress has been introduced and used within a local density approximation (LDA) framework to explain the orthorhombic distortion~\cite{Filippetti1999, Filippetti2000}. Recently it was shown that taking strong electron correlations into account in the calculations at the level of the local spin density approximation plus a Hubbard U-term (LDA+U) could improve the agreement between calculations and experiments by opening up a small band gap at the Fermi level~\cite{Herwadkar2009}. However, all these calculations considered only ordered magnetic structures while most experimental measurements, especially of the band structure, are performed above the N\'eel temperature.

Unlike what is sometimes assumed, most magnetic systems retain magnetic moments also above their critical Curie or N\'eel temperature. Indeed local moments are typically present although long-range order between them is lost. CrN is such a system where the experimentally observed structural (lattice spacing) and electronic properties (semiconducting behavior) of the paramagnetic cubic phase can not be even qualitatively reproduced by non-magnetic calculations~\cite{Filippetti1999}. At the same time, when performing first-principles calculations modelling such disordered cases, ordered magnetic structures should not be used because they might give rise to order-specific features, like the well known mixing anomaly in the Fe$_{1-x}$Cr$_x$~\cite{Olsson2003} system. This means that a disordered magnetic state must be considered in order to fully understand the physics of paramagnetic CrN at room temperature.

 For such a purpose the disordered local moments (DLM)~\cite{Gyorffy1985} method has been suggested and implemented within the coherent potential approximation (CPA)~\cite{Soven1967} treatment for disorder. The CPA-DLM method was used in Ref.~\cite{Alling2007JAP} to demonstrate the importance of the magnetic degree of freedom when paramagnetic cubic CrN was alloyed with AlN. Even though the DLM-CPA treatment of magnetic disorder is an excellent approximation in many cases, a direct method for calculating the electronic structure of magnetically disordered systems within a conventional supercell methodology is highly desirable. This is so since the CPA is most often combined with other approximations. e.g. the spherical approximation for the single-particle potential, imposing certain limitations on the treatment of materials with complex underlying crystal lattices. Moreover, in magnetic alloys, such as Cr$_{1-x}$Al$_{x}$N or Fe$_{1-x}$Ni$_{x}$~\cite{Abrikosov2007, Liot2009}, or in the presence of defects such as nitrogen vacancies in CrN, local lattice relaxations and other local environment effects might be important and they are beyond the reach of the single-site CPA theory. Furthermore, if the magnetic and vibrational thermodynamics of solids are ever to be treated simultaneously on the same footing, a supercell treatment of magnetism, compatible with quantum molecular dynamics simulations needs to be developed. 
 
In this work we apply two different supercell approaches to treat the magnetic disorder of the paramagnetic phase of CrN and compare them with CPA-DLM calculations. Firstly we apply the special quasirandom structure (SQS) method~\cite{Zunger1990}, developed to treat chemically disordered alloy systems. Secondly, to gain further confidence, we propose a magnetic sampling method (MSM) and show that the two methods give equivalent results. 

Furthermore we investigate the impact of strong electron correlations on structural and electronic properties of CrN at the level of LDA+U calculations. Considering both magnetic disorder and strong electron correlations simultaneously we analyze the magneto-structural transition in CrN.

\section{Calculational details}
In this work electronic structure calculations are performed within a density functional theory (DFT) framework and the projector augmented wave (PAW) method~\cite{Blochl1994} as implemented in the Vienna \emph{ab-initio} simulation package (VASP)~\cite{Kresse1993, Kresse1999}. Both the local spin density approximation (LDA)~\cite{Ceperley1980}, the generalized gradient approximation (GGA)~\cite{Perdew1996}, and a combination of the LDA with a Hubbard Coulomb term (LDA+U)~\cite{Anisimov1991,Dudarev1998} methods are used for treating electron exchange-correlation effects. The Hubbard term is applied only to the Cr 3d orbitals. In this implementation of the LDA+U method there is only one free parameter corresponding to U$^{eff}$=(U-J). In the following the simple notation U is used for U$^{eff}$. The energy cut-off for plane waves included in the expansion of wave-functions are 400 eV. Sampling of the Brillouin zone was done using a Monkhorst-Pack scheme~\cite{Monkhorst1972} on a grid of 5x5x5 (64-atom supercells), 9x9x9 (48-atoms supercells), 13x13x13 (8-atom cells), and 21x21x21 (2-atoms cells) k-points. To find the optimal cell geometry for the orthorhombic structure an automatic optimization procedure was used independently for each volume. We also apply the exact muffin-tin orbitals (EMTO) method~\cite{Vitos1997, Vitos2001PRL} including the full charge-density technique~\cite{Vitos2001} in which the CPA-DLM treatment of magnetic disorder is implemented. The EMTO basis set included s, p, d, and f-orbitals and the total energies were converged within 0.5 meV/f.u. with respect to the density of the k-point mesh. 


\section{Supercell approach to magnetic disorder}
\subsection{The energy of a disordered magnet}
In this work we discuss the thermodynamics of the high temperature paramagnetic state of CrN within the classical Heisenberg description, known to work well for many itinerant magnetic systems. This state can be described as a disordered distribution of Cr magnetic moments on the lattice, lacking long-range order. Similar situations are believed to be present in many other systems such as bcc Fe and its alloys at high temperatures~\cite{Heine1990, Kissavos2006}, NiMnSb above the Curie temperature~\cite{Alling2009} as well as a large number of f-electron systems~\cite{Niklasson2003}. The first problem we encounter is thus how to calculate the energy of such a disordered magnet using a supercell technique. In order to create an adequate supercell that can be used we need to know the characteristics of the random distribution. An ideally random distribution of magnetic spins, corresponding in the Heisenberg model to infinite temperature, is characterised by the vanishing of all the average spin correlation functions

\begin{equation}\label{eq:Disorder}
<\Phi_\alpha>=\frac{1}{N}\sum_{i,j\subset \alpha}\mathbf{e}_i \cdot \mathbf{e}_j = 0,  \forall \alpha,
\end{equation}
where $\alpha$ corresponds to a specific coordination shell, $\mathbf{e}_i$ is a unit vector in the direction of the magnetic moment on site $i$ ($\mathbf{e}_i = \mathbf{S}_i/M$ where $M$ is the magnitude of the magnetic moments) and N is a normalisation constant. If the magnetic properties are approximated within a classical Heisenberg model hamiltonian, we get

\begin{equation}\label{eq:Heisenberg}
H_{mag}=-\sum_{i\neq j}J_{ij}\mathbf{e}_i\cdot \mathbf{e}_j = -\sum_{\alpha}J_{\alpha}n_{\alpha}<\Phi_{\alpha}>,
\end{equation}
where $n_{\alpha}$ is the number of atoms in the $\alpha$:th coordination shell on the lattice.

The magnetic moments in CrN have been shown to be formed due to a magnetic split of the Cr t$_{2_g}$ non-bonding d-states present at the Fermi level~\cite{Filippetti1999}. Since large local Cr-moments are present and showing rather stable values regardless if they are ordered in a ferromagnetic configuration, antiferromagnetic configuration or a disordered local moments configuration~\cite{Alling2007JAP}, it seems reasonable that a Heisenberg model hamiltonian could describe the magnetism of CrN rather well. We note that this model, although limited to the first two nearest neighbour interactions, was applied in Ref.~\cite{Filippetti2000} for analyzing magnetic induced stress. However, one should be aware that it has been shown that the interaction parameters $J_{ij}$ could depend quite substantially on the global magnetic state even if the magnetic moments are almost constant~\cite{Alling2009}, underlining the importance of a reliable method directly accessing the disordered magnetic state. 

Two important observations can be made from Eq.~2. Firstly, in order to calculate the total energy of the ideal random distribution, the structure we use in simulations has to fulfill the condition in Eq.~1 only for the coordination shells $\alpha$ where the interaction parameters $J_{\alpha}$ are non-negligible. Secondly, one realizes that even though the disordered spin state is utterly non-collinear, its energy can be calculated using a collinear state as long as the parallel ($\mathbf{S}_i \cdot \mathbf{S}_j = +1\cdot M^2$) and anti-parallel ($\mathbf{S}_i \cdot \mathbf{S}_j = -1\cdot M^2$) spin pairs exactly cancel on each relevant coordination shell. Thus, it is these characteristics that we should aim for in our simulation of the high temperature paramagnetic state.

It is not directly obvious how to create a supercell fulfilling these properties but it is in fact a situation very similar to the problem of modeling chemical disorder in the form of a binary substitutional random alloy~\cite{Ruban2008REV}. The DLM-CPA method mentioned above can actually be seen as just a CPA treatment of a random alloy of atoms with spin up and spin down oriented magnetic moments. If we follow this analogy to the supercell framework it is the special quasirandom structure (SQS) methodology, first suggested by Zunger~\emph{et al.}~\cite{Zunger1990}, that has proven to be the most accurate approach for direct calculations of the total energy or related properties of the disordered state. The agreement between the CPA and SQS methods for chemical disorder in the B1 structure was demonstrated for the Ti$_{1-x}$Al$_x$N system in Ref.~\cite{Alling2007} using the same electronic structure methods as in this work. Here we use the 48-atom (24 metal + 24 nitrogen) SQS structure suggested in Ref.~\cite{Alling2007} for Ti$_{0.5}$Al$_{0.5}$N to model the Cr$_{0.5}^\uparrow$Cr$_{0.5}^{\downarrow}$N paramagnetic phase. That structure has $<\Phi_{\alpha}>=0$ for the first 7 coordination shells with the exception of a small non-zero value on the fifth shell. For comparison reasons we also use a 64-atom (32 chromium + 32 nitrogen) SQS-structure based on a cubic 2x2x2 conventional unit cell geometry  with $<\Phi_{\alpha}>=0$ for the first 6 shells with the exception of a small non-zero value on the third shell. We used the SQS method in a study of the bulk modulus of the paramagnetic phase of CrN~\cite{Alling2010natmat} as well as to get a CrN reference energy in a study of Ti$_{1-x}$Cr$_x$N~\cite{Alling2010TiCrN}. However the reliability of the method was questioned in Ref.~\cite{Rivadulla_answer}.

To check if the SQS method is reliable to model magnetic disorder on the cubic lattice of CrN we show in Fig.~1 a comparison of energy-lattice parameter curves for different magnetic states including the disordered state calculated with the SQS-PAW (left panel) and CPA-EMTO (right panel) methods employing the GGA functional for exchange-correlation effects.  The energies are given relative to the non-magnetic energy minimum. The agreement between the two different treatments of disordered magnetism is clearly seen as the disordered state is placed in a very similar relation to ordered magnetic and non-magnetic calculations in the two methodological frameworks. 

\begin{figure}
\includegraphics[angle=-90,width=0.87\columnwidth]{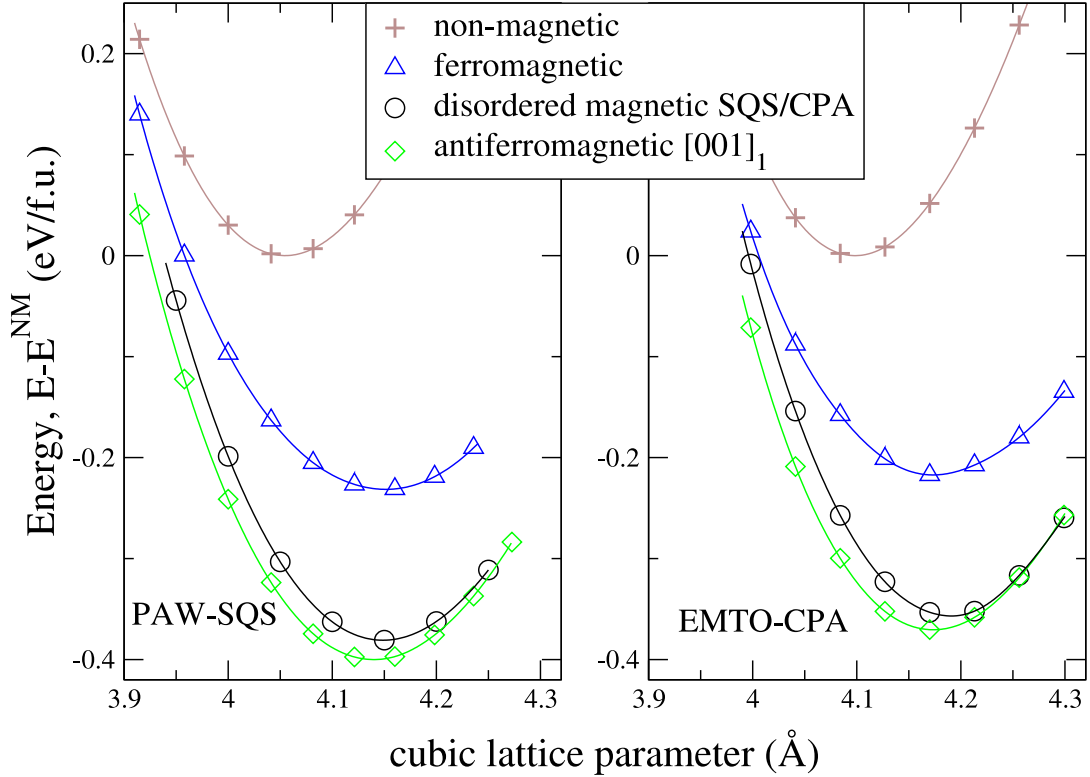}
\caption{\label{fig:EMTO_PAW} (Color online) The energies as a function of lattice parameter of different magnetic states of cubic B1 CrN as calculated with the PAW and EMTO methods and the GGA exchange-correaltion functional. The disordered phase (circles) is modelled with the SQS method in the PAW calculations and the DLM-CPA method in the EMTO calculations.}
\end{figure}

Here we note that the accuracy of the DLM-CPA treatment of a completely disordered magnetic state is established analytically on the single-site level in Ref.~\cite{Gyorffy1985}. We therefore view very good agreement between SQS and DLM-CPA calculations as a strong proof that the former technique is capable to describe the energy of a paramagnetic state, at least at the same level of accuracy. In particular, it is clear that the SQS method does not suffer from imposed periodic boundary conditions or the fact that one technically speaking deals with one selected antiferromagnetic configuration. As soon as conditions given by Eqs.~\ref{eq:Disorder} and \ref{eq:Heisenberg} are fulfilled, the SQS represents a quasi-random rather than ordered magnetic state. As for some minor differences between the results presented in the two panels of Fig.~\ref{fig:EMTO_PAW}, they come from the usage of different underlying methods for the electronic structure calculations, PAW and EMTO.

In order to further establish the reliability of the SQS approach, using the same PAW methodology, we first compare the results calculated for the two different SQS geometries considered in this work. The energy difference between them is 0.003 eV/f.u. Due to the translational symmetry the SQS based on the 2x2x2 conventional unit-cells has the problem that the correlation function on the 8:th correlation shell is exactly 1 and this is probably the main source of the small difference between the two SQSs. 

Next we suggest a different method to calculate the energy of a magnetic state approximating that in Eq.~1.: The magnetic sampling method (MSM). Within the MSM the directions (up and down) of magnetic moments of the Cr-sites of a large supercell are chosen using a random number generator. A large set of different such distributions, magnetic samples, are then created. Their energies are calculated and the \emph{average energy} is taken as the energy of the disordered state. Individually these supercells typically do not satisfy the conditions of Eqs.~\ref{eq:Disorder} and~\ref{eq:Heisenberg} but their average should, given that a sufficiently large number is considered. Although the SQS formalism was suggested in a reaction against the inaccuracies of random number distribution schemes~\cite{Zunger1990}, we note that the supercell sizes and particularly the number of calculations possible to treat with todays computational resources are orders of magnitude larger as compared to those back in 1990.


\begin{figure}
\includegraphics[angle=-90,width=0.87\columnwidth]{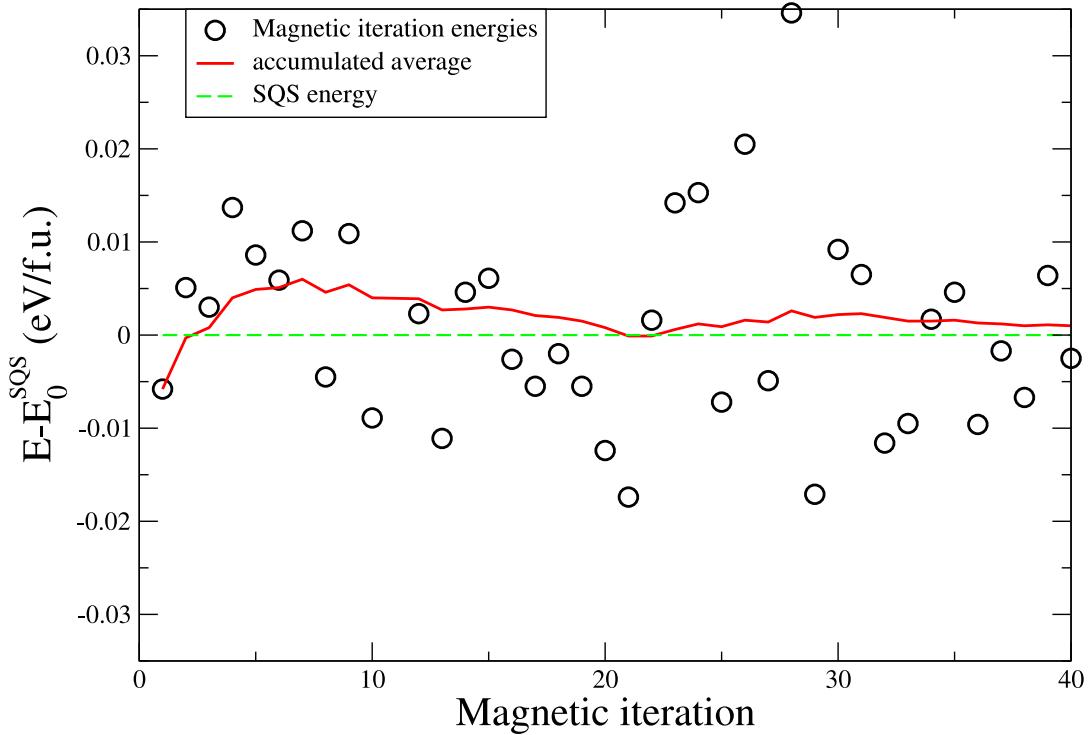}
\caption{\label{fig:0_ENERGY} (Color online) The calculated energies of the different random number generated magnetic configurations of a 32 Cr-atoms B1 CrN supercell (circles). The accumulated average of the random number generated configurations, called magnetic iterations, is shown with a solid line. All values are plotted relative the energy minimum of a SQS based magnetic configuration on the same underlying structure.}
\end{figure}

Fig.~\ref{fig:0_ENERGY} shows the calculated energies of 40 different randomly generated magnetic samples with up and down collinear moments on the ideal lattice points of a 32 Cr-atoms (2x2x2 conventional unit cells) B1 CrN supercell. Their accumulated average energy is shown with a solid line and compared to the energy of the SQS generated configuration on the same underlying geometry. The latter is taken as the reference energy. Although the individual energies of the randomly generated supercells can differ as much as -0.02 eV/f.u. and +0.035 eV/f.u. from the SQS-value, the accumulated average after the consideration of 40 different samples, called magnetic iterations, differs only by 0.001 eV/f.u. Between iteration 20 and 40, the average is never more than 0.001 eV/f.u. above or 0.0015 eV/f.u. below the value at iteration 40 showing that the mean energy is converging. If needed one could use even more sampled structures to converge the value with a higher accuracy than the 40 used in the present work. With this said we conclude, by our comparison with the CPA-DLM calculations as well as the internal agreement between SQS and MSM methods that both the considered supercell approaches can be used to calculate the total energy of a disordered collinear magnetic state of CrN with an accuracy of a few meV/f.u. on a fixed ideal B1 lattice.

\subsection{Non-collinear considerations}
Finally, we note that for systems where the energetics of the non-collinear disordered state is believed not to be well described by Eq.~\ref{eq:Heisenberg}, for instance due to non-negligible contributions from bi-quadratic terms in the Hamiltonian:

\begin{equation}\label{eq:Q_Heisenberg}
H_{mag}=-\sum_{i\neq j}J_{ij}\mathbf{e}_i\cdot \mathbf{e}_j -\sum_{i\neq j}K_{ij}\left( \mathbf{e}_i\cdot \mathbf{e}_j \right)^2= -\sum_{\alpha} \left( J_{\alpha}n_{\alpha}<\Phi_{\alpha}>+K_{\alpha}n_{\alpha}<\Psi_{\alpha}>\right) ,
\end{equation}
where in the fully disordered state
\begin{equation}\label{eq:Q_Disorder}
<\Psi_\alpha>=\frac{1}{N}\sum_{i,j\subset \alpha}\left( \mathbf{e}_i \cdot \mathbf{e}_j \right) ^2 = \frac{1}{3}, ~\forall~\alpha,
\end{equation}

\noindent the MSM method could still be used with a straight forward generalisation: Instead of random number generation of collinear spins (up and down in {\bf \^z}), one can generate a set of different non-collinear supercells with six types of local moments describing up and down along {\bf \^x}, {\bf \^y}, and {\bf \^z}. Such a set, given that the supercells are large enough and that the number of supercells is large enough, will reproduce both the bi-linear (Eq.~\ref{eq:Disorder}), and bi-quadratic (Eq.~\ref{eq:Q_Disorder}) correlation functions of the disordered state. In principle, but cumbersome in practise, also the SQS method can be used in this case by constructing a large supercell of 6 components with vanishing correlation functions between them all. Unfortunately, first-principles calculations of non-collinear magnetic systems are rather time consuming, nevertheless we have performed non-collinear MSM calculations within the LDA+U approximation for the volume obtained with the collinear approximation and will get back to this issue below. 



\section{Effect of strong electron correlations}
Having established methods to treat the magnetic disorder within a supercell framework we now turn to the problem of electron exchange-correlation energies. Even though LDA calculations qualitatively revealed the energetics of CrN~\cite{Filippetti1999, Filippetti2000} the electronic structure did not reproduce the experimentally observed semiconducting behavior. Thus one could doubt the accuracy of LDA predictions of structural and magnetic energy differences of relevance for understanding the orthorhombic to cubic transition in this system. Recently Herwadkar~\emph{et al.} studied CrN using the LDA+U approach with focus on the electronic structure of ordered magnetic structures. They calculated the value of U and J, the screened Coulomb and exchange terms respectively, using constrained LDA approaches and achieved U=3 eV and J=0.9 eV~\cite{Herwadkar2009}. However, they suggested a span of U values from 3-5 eV to be reasonable. Even though such an \emph{ab-inito} approach to obtain the values of the U and J parameters are appealing, the uncertainties are to large for a quantitative a thermodynamic analysis.

In order to obtain the most suitable value $U^{eff}$, for which the LDA+U method best describes the relevant properties of CrN, we perform a careful comparison of structural parameters and electronic structure obtained with LDA+U calculations for various values of U with experiments. For comparison also the results obtained with the generalized gradient approximation, GGA, are presented. 

\begin{figure}
\includegraphics[angle=-90,width=0.87\columnwidth]{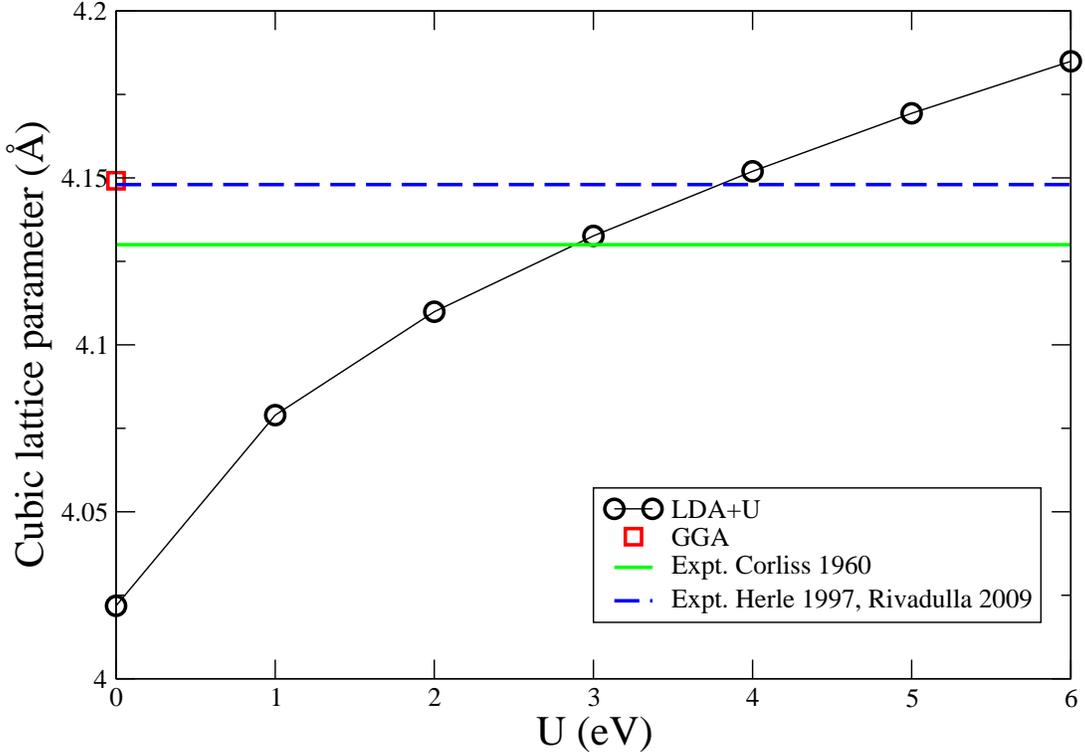}
\caption{\label{fig:latpar} (Color online) The calculated cubic lattice parameter for the magnitically disordered state within the LDA+U (circles) and GGA (square) approximations. The experimental bulk value found by Corliss~\emph{et al.}~\cite{Corliss1960} and the common value found by Herle~\emph{et al.}~\cite{Herle1997} and Rivadulla~\emph{et al.}~\cite{Rivadulla2009}, are shown with solid and dashed horizontal lines respectively. }
\end{figure}

First the lattice parameter of the cubic paramagnetic phase, modeled with the SQS approach, is presented in Fig.~\ref{fig:latpar}. The experimental value obtained for bulk CrN by Corliss~\emph{et al.} is 4.13~\AA~\cite{Corliss1960}, while both Herle~\emph{et al.}~\cite{Herle1997} and, more recently, Rivadulla~\emph{et al.}~\cite{Rivadulla2009} obtained 4.148~\AA. Values obtained for CrN in thin films are typically slightly larger~\cite{Gall2002, Sanjines2002} but include strain effects not considered in the calculations. The result obtained with the pure LDA functional is a=4.022~\AA. Using GGA we obtain a=4.149~\AA. The LDA+U approach gives increasing lattice spacing with increasing U. The lattice parameter of Corliss~\emph{et al.} is obtained with U=2.9 eV. The value of Herle~\emph{et al.} and Rivadulla~\emph{et al.} is obtained with U=3.8 eV. Since the reported experimental values are measured at room temperature while the calculations, with the exception for the magnetic disorder, corresponds to a 0~K situation, one might object that the comparison is not completely fair. However, thermal expansion between 0~K and room temperature is typically small in this class of hard ceramics. 

\begin{figure}
\includegraphics[bb=  56 79 537 761,angle=90,width=0.87\columnwidth]{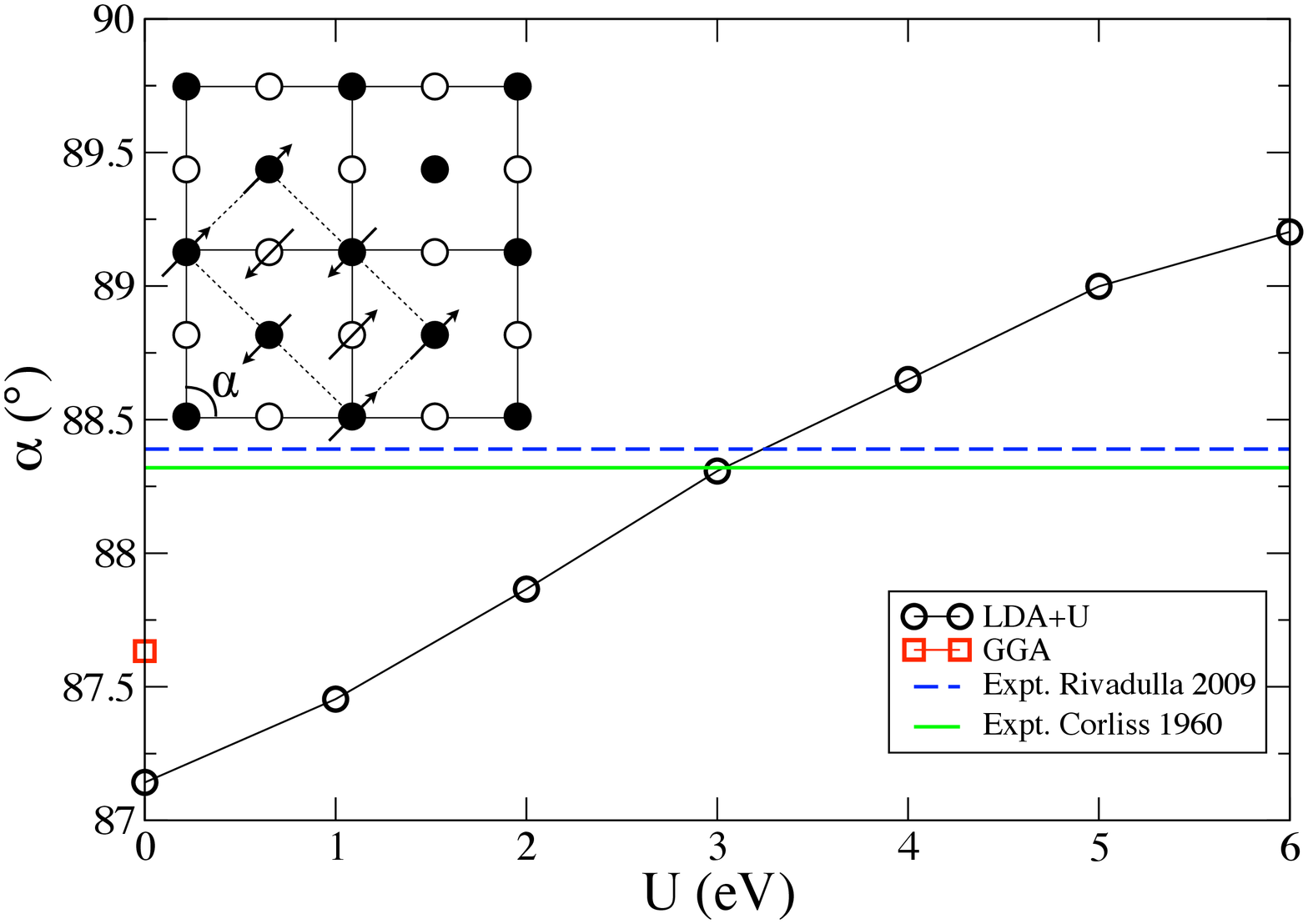}
\caption{\label{fig:alpha} (Color online) The calculated angle $\alpha$ in the distorted orthorhombic state between the axis of the conventional B1-cell. Values using the LDA+U approximation (circles) and the GGA approximation (square) are shown together with the experimental value found by Corliss~\emph{et al.}~\cite{Corliss1960} (solid horizontal line) and Rivadulla~\emph{et al.}~\cite{Rivadulla2009} (dashed horizontal line). Inset: The definition of the angle $\alpha$ with respect to the orthorhombic unit cell (dashed lines) shown in a 001-plane of the cubic B1 structure. The relative directions of the spins of Cr atoms following Ref.~\cite{Corliss1960} are also shown (solid circles: z=0.0, open circles: z=0.5).}
\end{figure}

Since also the physics of the orthorhombic phase must be well described by our theoretical model, we compare in Fig.~\ref{fig:alpha} the calculated value of the angle $\alpha$ with the experimental finding in Refs.~\cite{Corliss1960, Rivadulla2009}. $\alpha$ describes the angle between the axis of the conventional unit cell of the cubic B1 lattice, that is distorted in the orthorhombic phase, see the inset in Fig.~\ref{fig:alpha}. The experimental value is 88.3$^\circ$according to Corliss~\emph{et al.}~\cite{Corliss1960} or 88.4$^\circ$~according to Rivadulla \emph{et al.}~\cite{Rivadulla2009}, while both LDA and GGA calculations underestimate this angle, thus overestimating the distortion. On the other hand, as can be seen in Fig.~\ref{fig:alpha}, a value in agreement with the experiment is obtained with the LDA+U method for U=3.0 eV.

\begin{figure}
\includegraphics[angle=-90,width=0.83\columnwidth]{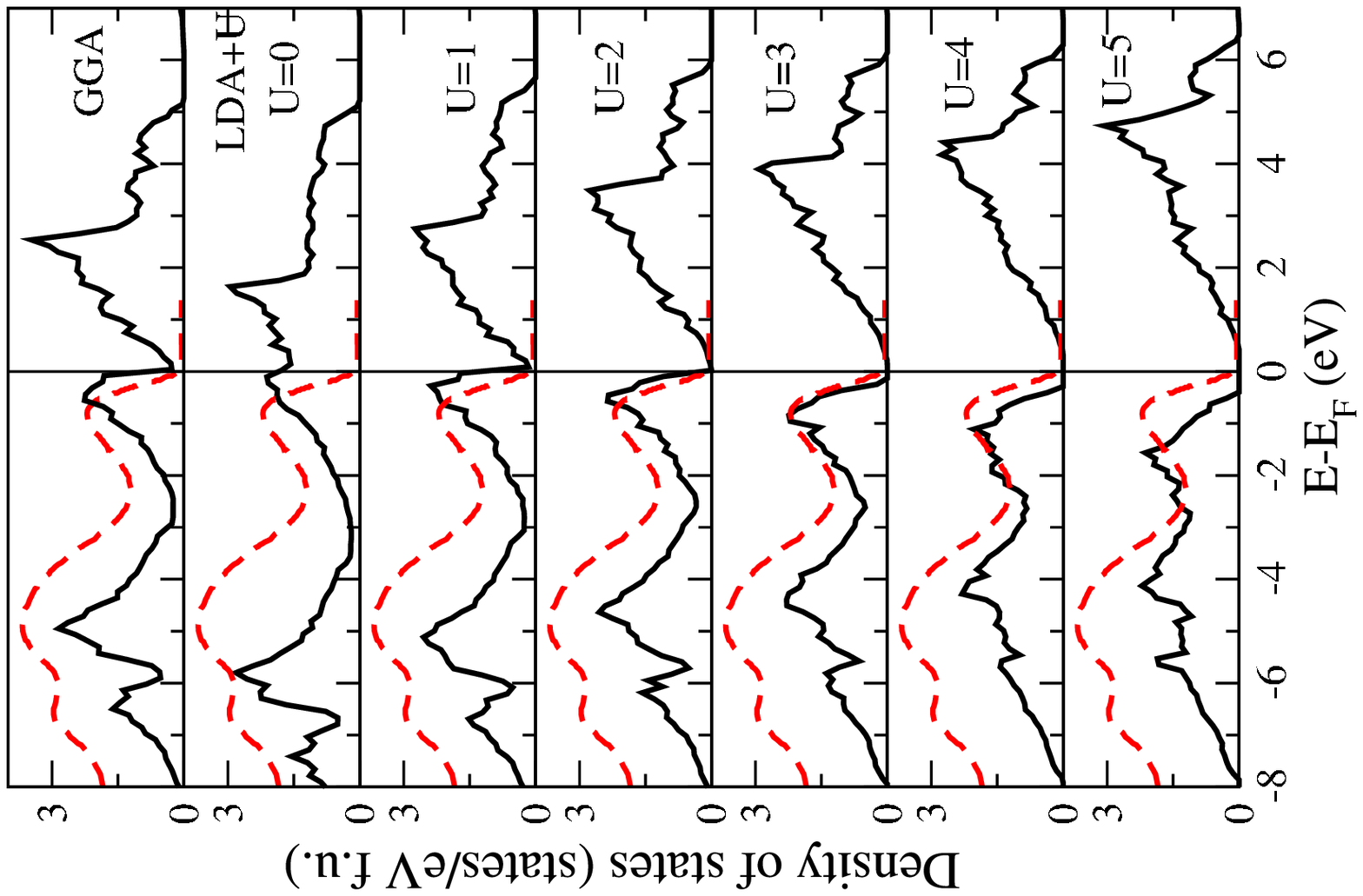}
\caption{\label{fig:dlmU} (Color online) The calculated valence band electronic density of states (solid line in all frames) of the cubic magnetically disordered state using the GGA approximation (top panel) and LDA+U approximation with different values of U. For comparison the experimental UPS spectroscopy measurement by Gall~\emph{et al.}~\cite{Gall2002} is shown by a dashed line in all frames.}
\end{figure}

Finally, in Fig.~\ref{fig:dlmU} we compare the calculated total electronic density of states of the valence band of the disordered magnetic state (calculated with the SQS method) with the ultraviolet photoemission spectroscopy measurement of the cubic paramagnetic phase obtained by Gall~\emph{et al.}~\cite{Gall2002}. Figure~\ref{fig:dlmU} shows in different panels (from top to bottom) the DOS obtained with GGA, and LDA+U with U from 0 eV (LDA) to U=5 eV. In all panels the experimental results are shown with dashed lines. In all cases we use the lattice spacing corresponding to the equilibrium of the particular choice of exchange-correlation scheme. The GGA, and even more so the LDA, gives an overlap of the peaks close to the Fermi level. These peaks correspond primary to Cr spin-up non-bonding (below $E_F$) and a combination of spin down Cr non-bonding and Cr spin up anti-bonding states (above $E_F$). Both peaks also have small ad-mixture of N p-character~\cite{Herwadkar2009}. When the U value in the LDA+U approach is increased, the occupied Cr spin up non-bonding state becomes more localized and shift down in energy. At the same time the unoccupied states are shifted up and at a value of U between 2 and 3 eV a small gap opens at $E_F$, in agreement with the experiment~\cite{Gall2002}. Actually the LDA+U approximation with U=3~eV excellently describes the Cr spin up non-bonding state, while the bonding states on the other hand are shifted to slightly too high energies. 

In summary, the LDA+U approximation with U-values between about 3 and 4 eV reproduces the cubic paramagnetic lattice parameter, the value U=3 eV reproduces experimental measurement of the angle $\alpha$ in the orthorhombic phase, and a value of U between 2 and 3 eV gives a good description of the electronic structure of the valence band of the cubic paramagnetic phase. Thus we conclude that within the LDA+U approximation, the value U$^{eff}$=3 eV gives an optimal description of the physical properties of the system. This value is safely within the range suggested in Ref.~\cite{Herwadkar2009} corresponding to U=3.9 if their choice of J=0.9 is used.

In the following sections we use the LDA+U approximation with effective U=3 eV in our calculations. However, since the GGA approximation has already been used in many studies on related systems, we present also results using the GGA to illustrate the effect of strong electron correlations.

\section{{Energetics of magnetic and crystallographic phases of C\MakeLowercase{r}N}}
\begin{figure}
\includegraphics[angle=-90,width=0.83\columnwidth]{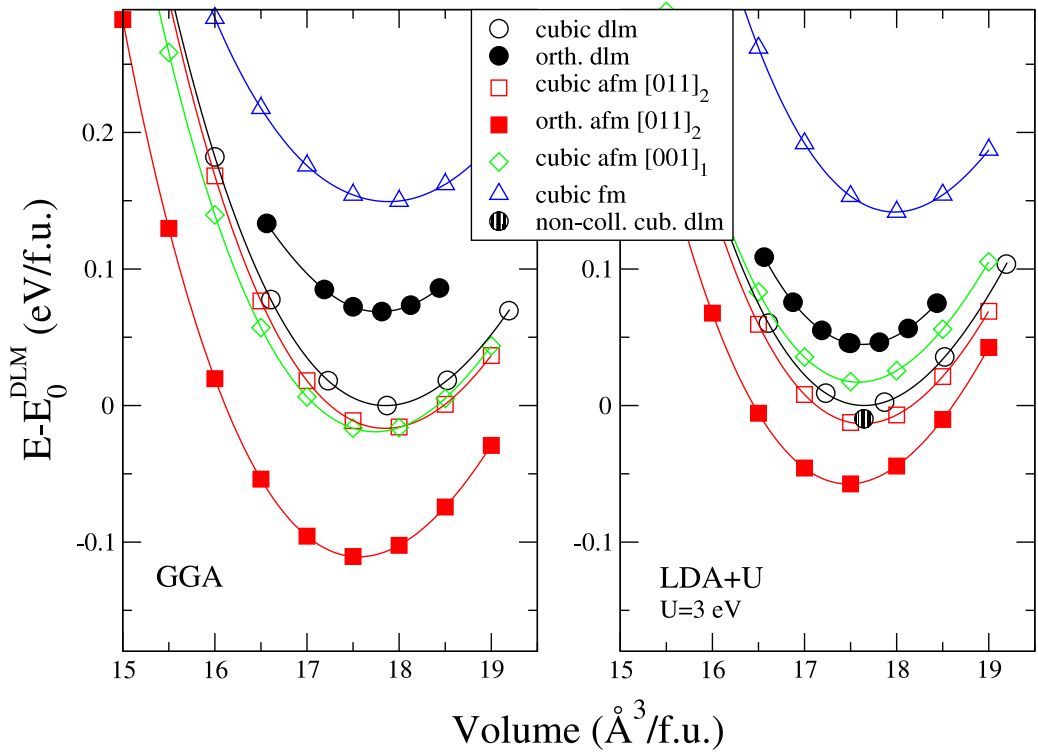}
\caption{\label{fig:EOS} (Color online) The calculated energy versus volume curves for different magnetic states in CrN using the GGA approximation (left panel) and LDA+U (U=3eV) (right panel). The energy minimum of the cubic collinear disordered local moment state (DLM) is used as reference energy. }
\end{figure}

We now have the theoretical tools needed to study the magneto-structural transition in CrN. In Fig.~\ref{fig:EOS}~we consider the total energies, as a function of volume of different crystallographic and magnetic phases of CrN. The left panel shows the results obtained with the GGA functional while the right panel shows the results obtained with LDA+U (U=3 eV). Cubic phases are shown with open symbols while orthorhombic (orth.) structures are shown with solid symbols. The paramagnetic cubic phase is modeled with the SQS disordered local moments method (denoted cubic dlm) keeping the atoms fixed at B1 lattice points. The energy of a disordered magnetic configuration on the lattice points of the orthorhombic structure is also shown for comparison and denoted orth. dlm. The minimum energy of the cubic collinear disordered magnetic phase is taken as the reference value. Furthermore, the energy of the experimental ground state structure~\cite{Corliss1960}, the orthorhombic distorted double [011]-layered antiferromagnetic ([011]$_2$ afm) structure schematically shown in the inset of Fig.~\ref{fig:alpha}, is shown as is the energy of the same magnetic ordering on the cubic lattice. Also a single layer [001]-ordered antiferromagnetic state ([001]$_1$ afm) on the cubic lattice, the ferromagnetic cubic phase, and in the case of LDA+U the energy of the non-collinear cubic magnetically disordered phase are shown. 

One can see that the GGA and LDA+U calculations are in reasonable qualitative agreement with each other. However, the GGA gives considerably larger energy differences between the orthorhombic and cubic phases as compared to the LDA+U calculations. Also the order of the two considered antiferromagnetic states for the cubic phase are reversed. In the GGA framework the $[001]_1$ antiferromagnetic state is lower in energy, in line with previous works~\cite{Filippetti1999, Alling2007JAP}.  In the LDA+U framework on the other hand, the $[011]_2$ antiferromagnetic state is lowest in energy also in the cubic phase.

One interesting comparison can be made between the introduction of a Hubbard U-term in this work and the alloying of CrN with AlN studied in Ref.~\cite{Alling2007JAP}. In that work it was found that the DLM-CPA state became lower in energy as compared to the $[001]_1$ antiferromagnetic state when a certain amount of Al was substituted for Cr. The point is that upon alloying of CrN with AlN~\cite{Alling2007JAP}, and more generally upon alloying transition metal nitrides with AlN~\cite{Alling2008Surf}, the inclusion of Al favours a localisation of the transition metal non-bonding d-states. Obviously, the strong electron correlations leads to a similar effect which explains the similar evulotion of the magnetic energies in the two cases.

\section{Non-collinear magnetism in C\MakeLowercase{r}N}
Using the LDA+U approximation the impact of non-collinear magnetism on the energy of the disordered cubic phase has been investigated with the MSM method by averaging over 35 different random-number generated non-collinear distributions, a procedure capable of reproducing the condition of both Eqs.~\ref{eq:Disorder} and~\ref{eq:Q_Disorder} for the relevant coordination shells. The average energy is found to converge in a similar manner as the collinear case described above. The energy of the non-collinear disordered magnetic state is found to be 0.010 eV/f.u. below that of the collinear disordered state and is marked in Fig~\ref{fig:EOS} with a striped circle. This energy difference is not particularly large with respect to e.g. the orthorhombic antiferromagnetic state or the cubic ferromagnetic state, but as will be seen below, do have a quantitative effect on the simulation of the cubic-to-orthorhombic transition temperature. In our case the improved level of theoretical modelling when including non-collinear effects, came to the cost of an approximately seven-fold increase in computational time with respect to otherwise identical settings for the calculations of collinear disordered structures.

\section{The magneto-structural transition in C\MakeLowercase{r}N}
Qualitatively our calculated values agree with the experimental observation of the stability of the orthorhombic state at low temperatures and a cubic state at higher temperatures. In this work we denote the temperature for this structural transition $T_S$ in order not to confuse it with the hypothetical isostructural N\'eel temperatures, $T_N$, of a magnetic order-disorder transiton on a fixed lattice. This is so since the disordered paramagnetic phase has a considerable magnetic entropy making it more competitive at higher temperatures. Since the energy of the cubic dlm-state is considerably lower than the orthorhombic dlm-state the magnetic disordering is accompanied by a structural transition. The fact that no signs of magnetic ordering was observed in the epitaxially stabilized cubic phases in Ref.~\cite{Gall2002, Sanjines2002} can be understood from the fact the the energy differences between antiferromagnetic and disordered magnetic phases in the cubic geometry are small, especially if non-collinearity is taken into account, indicating a very low N\'eel temperature for undistorted cubic CrN. In the orthorhombic structure on the other hand the difference is almost an order of magnitude larger indicating that the orthorhombic antiferromagnetic state should be well below its isostructural N\'eel temperature at the experimental transition point $T_S^{expt}=280-287 K$~\cite{Corliss1960,Rivadulla2009}. This result is in line with the experimental observation that the extent of the orthorhombic distortion, measured with the value of the angle $\alpha$, is in principle the same at 286~K: $88.4~\degree$~\cite{Rivadulla2009}, 273~K: $88.3~\degree$~\cite{Browne1970}, and 77~K: $88.3~\degree$~\cite{Corliss1960} where the authors of the latter reference stated that no differences was seen when the temperature where further decreased down to the liquid Helium regime. If there had been a large degree of partial magnetic disorder in the orthorhombic phase one would expect a change in the value of this angle. Thus it is reasonable to assume that the transition is governed by the competition in terms of free energy between a disordered paramagnetic cubic phase with high magnetic entropy and a highly ordered antiferromagnetic orthorhombic phase with low magnetic entropy. Such a phase transition, including an abrupt change in both energy and entropy is in line with the experimental finding of a first order phase transition displaying a hysteresis behavior during heating-cooling cycles~\cite{Browne1970}.


Using our obtained structural energy differences we can estimate the transition temperature theoretically. At temperatures considerably above the (isostructural) N\'eel temperature, such as in the case of paramagnetic CrN at room temperature, the entropy of a Heisenberg system is well described by the mean-field term

\begin{equation}\label{eq:MF}
S^{mf}=k_{B}ln(M+1) ,
\end{equation}
where $M$ is the magnitude of the magnetic moment (in units of $\mu_B$) and $k_B$ is the Boltzmann constant. In the LDA+U approach we find that the average magnetic moments are $M^{LDA+U}=2.82 \mu_{B}$. In the GGA calculation $M^{GGA}=2.49 \mu_{B}$.

Using these values and the approximations above the transition temperature can be obtained from the condition that at the critical temperature, the two phases should have the same free energy, F:
\begin{equation}
F_{afm}^{orth}(T_S)=F_{para}^{cub}(T_S) \Longleftrightarrow E_{afm}^{orth}=E_{para}^{cub}-T_{S}S^{mf}
\end{equation}

where $T_S$ denotes the critical temperature for the structural transition which is in reality also the magnetic ordering temperature. In our case we get with the collinear approximation for magnetic disorder $^{col}T_S^{LDA+U}=498~K$ with the non-collinear calculations ${}^{nc}T_S^{LDA+U}=413~K$ and with the collinear GGA calculations $T_S^{GGA}=1030~K$. This mean-field estimates should be compared to the experimental value of $T_S^{expt}=280-287 K$~\cite{Corliss1960,Rivadulla2009}. It is well known that the mean field approximation could overestimate magnetic ordering temperatures by as much as 50\% as compared to more reliable thermodynamics treatments. However, in the GGA calculation the error is so large that we instead interpret this result as one more argument that the GGA approximation overemphasize the orthorhombic distortion, both in geometric distortions visible in Fig.~\ref{fig:alpha}, and in the energy differences between the orthorhombic and cubic phases. The transition temperature derived from the non-collinear LDA+U calculation is closer but still considerably above the experimental measurement, giving an overestimation of $T_S$ with 44\%. On the other hand, the large relative error may be due to the low absolute value of the transition temperature. The absolute error in $T_S$ is just a little bit above 100~K for LDA+U treatment including non-collinear effects, which is the most typical case for first-principles simulations of structural phase transitions. In particular we suggest that magnetic short-range order coupled with the vibrational degree of freedom is of importance to quantitatively determine the transition temperature in CrN. It should be noted that also nitrogen off-stoichiometry could influence the transition temperature.
Moreover, one should remember that the LDA+U approach is an approximate method not free from errors, for instance in the exact choice of U, possibly affecting quantitative values of the important structural energy difference. 

Recently Rivadulla~\emph{et al.} showed that the temperature induced orthorhombic to cubic phase transition could be reversed with increasing pressure~\cite{Rivadulla2009}. At room temperature, a pressure as low as 1-2 GPa was enough to push the system back into the orthorhombic structure. Magnetic measurements showed that it was the antiferromagnetic ordered structure that reappeared~\cite{Rivadulla2009}. Qualitatively the pressure effect can be understood from the results in Fig.~\ref{fig:EOS}: Since the orthorhombic phase is slightly lower in volume as compared to the paramagnetic cubic phase, it will be relatively more favorable at elevated pressures according to the minimization of the Gibb's Free energy

\begin{equation}\label{eq:Gibbs}
G(T,p)=E+pV-TS.
\end{equation}

\begin{figure}
\includegraphics[angle=-90,width=0.83\columnwidth]{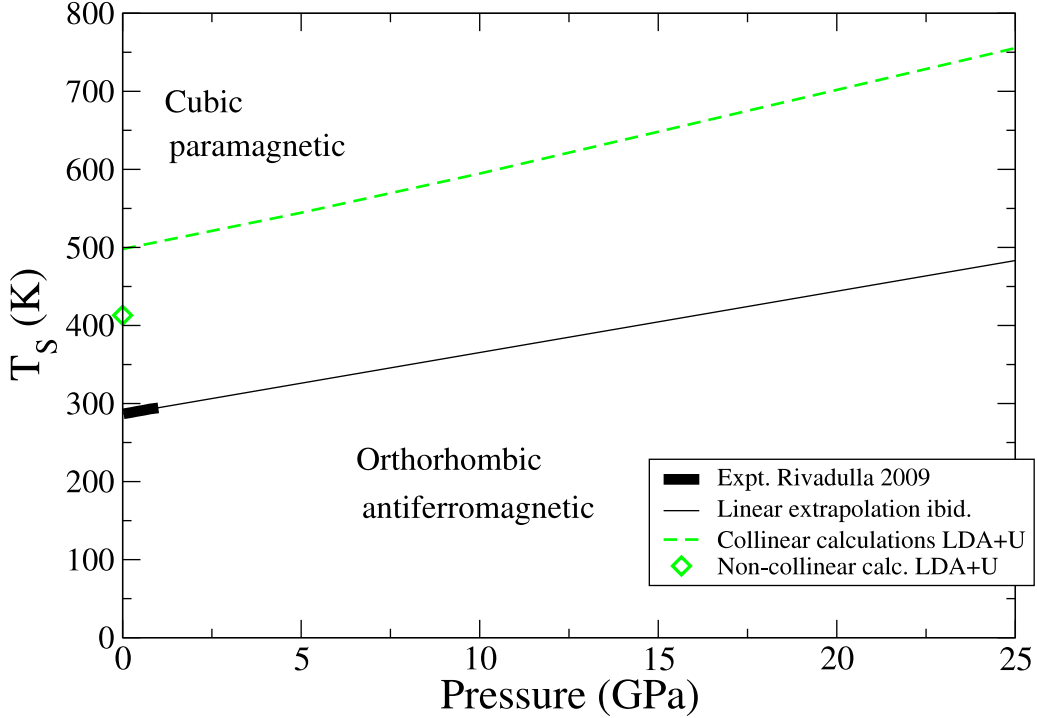}
\caption{\label{fig:PT_phase} (Color online) The calculated pressure-temperature phase diagram of CrN using the LDA+U (U=3eV) method and considering the disordered magnetic phase as collinear (green dashed line) or non-collinear (green diamond). The experimental measurements of Rivadulla~\emph{et al.}~\cite{Rivadulla2009} at low pressures are shown with a bold black line while a linear extrapolation to higher pressures is shown with a thin black line.}  
\end{figure}

The derived pressure-temperature phase diagram of CrN is shown in Fig.~\ref{fig:PT_phase}. The results from the theoretical calculations using Eq.~\ref{eq:Gibbs} with the mean field approximation for the magnetic entropy of the cubic phase, Eq.~\ref{eq:MF}, and the LDA+U approximation for exchange-correlation effects are compared to the experimental low-pressure results from Ref.~\cite{Rivadulla2009} and a linear extrapolation of these values to medium pressures. The qualitative picture, with increasing transition temperature with increasing pressure is rather well reproduced by the calculations although the absolute values of the temperatures are to high. Partially they are corrected by the consideration of non-collinear effects.

These results inspire us to propose a possible way to conclude the discussion of the value of the bulk modulus of cubic CrN~\cite{Rivadulla2009, Alling2010natmat,Rivadulla_answer}.  We suggest that in order to measure the compressibility of the cubic paramagnetic phase of CrN with higher accuracy as compared to Ref.~\cite{Rivadulla2009}, the experiment should be conducted at slightly higher temperatures where the cubic phase is stable over a larger pressure range. 

\section{Conclusions}
We have used two different supercell approaches to model disordered magnetism of paramagnetic materials, the special quasirandom structure method and the magnetic sampling method, and applied them for the study of CrN. The SQS method is valid and swiftly applicable when non-collinear effects can be neglected. The MSM and SQS mehods are shown to give equivalent results for collinear calculations of disordered local moments on a fixed B1 lattice in cubic CrN and both of them agree with DLM-CPA calculations. We show that it is straight-forward to extend the MSM method to calculations of non-collinear disordered magnetism. The latter is shown to have a quantitative influence on the calculated structure transition temperature in CrN.

CrN is a correlated material which is better described with a LDA+U aproach then with the GGA or LDA functionals. By comparing the calculated structural parameters and electronic structure of CrN with experiments we find that U$^{eff}$=3 eV is a suitable value to use in the simulations. 

Considering both magnetic disorder effects and strong electron correlations, the orthorhombic to cubic phase transition of CrN as a function of temperature and pressure can be qualitatively explained. In particular we show that it should be understood as a transition from a magnetically \emph{ordered} orthorhombic phase to a magnetically \emph{disordered} cubic phase. Considering magnetic entropy within the mean field approximation, the calculated transition temperature T$_S=413~K$ is a slight overestimation of the experimental value $280-287~K$. Magnetic short range order coupled with vibrational effects are likely to be of importance for determining the quantitative value of T$_S$. Since the transition also depends sensitively on the structural energy difference between the cubic and orthorhombic phases, which is shown to be very sensitive to the exchange-correlation functional, a strong electron correlations method beyond the LDA+U approach might be needed to reveal the details of the CrN phase transition.   


\section{Acknowledgement}
The Swedish Research Council (VR), the Swedish Foundation for Strategic Research (SSF), and the G\"oran Gustafsson Foundation for Research in Natural Sciences and Medicine are acknowledged for financial support.
Calculations were performed using computational resources allocated by the Swedish National Infrastructure for Computing (SNIC).

\end{document}